\documentclass{JHEP3}
\usepackage{epsfig}
\relax

\def\bea{\begin{eqnarray}}
\def\eea{\end{eqnarray}}
\def\nn{\nonumber}
\def\r{\rho}
\def\bes{\begin{eqnarray}}
\def\ees{\end{eqnarray}}
\def\be{\begin{equation}}
\def\ee{\end{equation}}
\def\bs{\begin{subequations}}
\def\es{\end{subequations}}
\newcommand{\een}{\end{subequations}}
\newcommand{\ben}{\begin{subequations}}
\newcommand{\beq}{\begin{eqalignno}}
\newcommand{\eeq}{\end{eqalignno}}

 \def\Ac{\mathcal{A}}

\def\rt{{\tilde{r}}}
\def\ti{{\tilde{t}}}
\def\phit{{\tilde{\phi}}}
\def\chit{{\tilde{\chi}}}

\def\be{\begin{equation}}
\def\ee{\end{equation}}
\def\ba{\begin{eqnarray}}
\def\ea{\end{eqnarray}}

\def\r{\rho}
\def\a{\alpha}

\def\e{\epsilon}

\def\m{\mu}

\def\s{\sigma}

\def\rt{\tilde r}

 \def\cT{{\cal T}}

\newcommand{\prt}[1]{{\left( {#1} \right)}}

\def\bs{\bigskip}

\def\IR{\relax{\rm I\kern-.18em R}}

\newcommand{\ff}{\frac}

\def\IR{\relax{\rm I\kern-.18em R}}
\def\IL{\relax{\rm I\kern-.18em L}}

\def\inv{^{\raise.15ex\hbox{${\scriptscriptstyle -}$}\kern-.05em 1}}

\def\bea{\begin{eqnarray}}
\def\eea{\end{eqnarray}}
\newcommand{\eq}[1]{(\ref{#1})}
\def\nn{\nonumber}

\newcommand{\la}[1]{\label{#1}}

\def\a{\alpha}

\def\e{\epsilon}

\def\m{\mu}

\def\r{\rho}
\def\s{\sigma}  
\def\t{\tau}

\title{Entropy of Thermal CFTs on Curved Backgrounds}

\author{D. Giataganas${}^{1,2}$ and N. Tetradis${}^{1}$
\\~\newline
${}^1$ Department of Physics, University of Athens,\newline 
Zographou 157 84, Greece\\
${}^2$ Institute of Advanced Study, Durham University,\newline
Cosin’s Hall, Palace Green, Durham, DH1 3RL

\\
{\tt dgiataganas@phys.uoa.gr, ntetrad@phys.uoa.gr}}

\preprint{}

\abstract{ We use holography in order to study the entropy of thermal CFTs on 
(1+1)-dimensional curved backgrounds that contain horizons. Starting from
the metric of the BTZ black hole, we perform explicit coordinate transformations
that set the boundary metric in de Sitter or black-hole form. For a de Sitter 
boundary, the dual picture
describes a CFT at a temperature different from that of the cosmological horizon.   
We determine minimal surfaces that allow us to compute the
entanglement entropy of a boundary region, as well as the temperature 
affecting the energy associated with a probe quark on the boundary. 
For an entangling surface that coincides with the horizon, we study the
relation between entanglement and gravitational entropy through an appropriate
definition of the effective Newton's constant.
We find that the leading contribution to the entropy is proportional to
the horizon area, with a coefficient that accounts for the degrees of 
freedom of a CFT thermalized above the horizon temperature. 
We demonstrate the universality of our findings by considering the most general 
metric in a (2+1)-dimensional AdS bulk containing a non-rotating black hole and a 
static boundary with horizons.
}


\begin{document}

\section{Introduction}
\label{intrr}

The relation between entanglement and gravitational entropy in 
spaces that contain horizons can shed light on the fundamental nature of 
gravity. The entanglement entropy, a non-local quantity measuring the correlation
between two subsystems, is UV divergent in continuum field theories. 
A crucial observation is that the divergent part 
scales with the area of the entangling surface \cite{sorkin}. 
The divergence is regulated through the presence of a physical cutoff in the theory.
When the entangling surface is identified with a horizon, the connection of
the entanglement entropy with the gravitational one must account for a relation
between the cutoff and Newton's constant. Such a connection may be possible 
if gravity is induced  by quantum fluctuations of matter fields \cite{jacobson}.

In recent work \cite{giataganas,correctionsds} we explored these issues in the
context of the AdS/CFT correspondence \cite{adscft1,adscft2}. 
The main difficulty one faces is that the 
the boundary metric is not dynamical, a feature
that is equivalent to the vanishing of the effective Newton's constant.
We based our analysis on the Ryu-Takayanagi conjecture \cite{ryu,review}, which
states that the entanglement entropy of a part of the 
boundary enclosed by an entangling surface $\Ac$ is proportional to the
area of a minimal surface $\gamma_A$ extending from $\Ac$ into the bulk. 
The area diverges near the boundary because of 
the short-distance entanglement of the local degrees of freedom 
on either side of the entangling surface,
which makes the introduction of a cutoff necessary. 
The proportionality factor involves the bulk Newton's constant.
Based on the approach of ref. \cite{hawking}, we determined the effective 
Newton's constant for the boundary theory by absorbing the cutoff in its
definition. For a (3+1)-dimensional boundary, our conceptual framework 
is equivalent to the Randall-Sundrum construction \cite{rs}, in which
dynamical gravity is generated by cutting off the AdS space before the
boundary is reached. 
In this context, the Newton's constant can be seen as
induced by quantum fluctuations of matter fields, represented by 
the bulk degrees of freedom that are integrated out.
The framework can be
extended to any number of dimensions by considering the regulated form of the
on-shell action in the context of holographic renormalization 
\cite{hol1,skenderis}.

Our approach differs from the usual interpretation, according to which 
the leading
contribution to the entanglement entropy is 
an unphysical UV-dependent quantity of little interest. On the contrary, it leads
to the opposite conclusion:
The leading contribution to the entropy has a universal
form that depends only on the horizon area because the same degrees
of freedom contribute to the entropy and Newton's constant.
The particular features of the underlying theory, such as the number of 
degrees of freedom, 
become apparent at the level of the subleading corrections to
the entropy \cite{correctionsds}.

In this work we apply our approach to new classes of boundary theories. We
focus on (1+1)-dimensional boundaries because of the formidable technical 
difficulties
that arise when dealing with higher-dimensional spaces. However, several
results can be obtained with interesting physical interpretation, which 
should be applicable to arbitrary dimensions.
The new element that we consider is the presence of a bulk horizon with
a characteristic temperature that may differ from the temperature 
of the boundary horizon. The simplest way to achieve this is by considering
a bulk black-hole metric with an arbitrary mass parameter. For a de Sitter boundary,
such constructions have been studied in ref. \cite{marolfds,fischlerds}.

In our setup the dual thermal field theories live in curved spacetimes. 
The bulk metric of the dual description 
can be thought of as emerging from appropriate slicings
of a bulk space that includes a black hole. 
This gravity dual has two types of horizons. For 
a de Sitter boundary, 
one is the cosmological horizon\footnote{When referring to the cosmological
horizon, we include its extension into the bulk.}, associated in general with the fact that a timelike geodesic observer sees a thermal bath of particles $T_{dS}=H/(2\pi)$.
The other is the bulk black-hole horizon that specifies 
the temperature of the dual field theory. While both of them indicate a thermal behavior, their crucial difference lies 
in the fact that the cosmological horizon is observer dependent and the 
effects of curvature cannot be easily disentangled from thermal effects. 
Our setup allows a special situation, where the bulk 
horizon has a different temperature than the horizon on the boundary. 
When the two temperatures coincide the dual field theory is in the canonical Euclidean Bunch-Davies vacuum, which reduces to the standard Minkowski one as we take the limit $H\rightarrow 0$. 
It seems counterintuitive that an equilibrium configuration exists with the
dual conformal field theory (CFT) at a different temperature than the one characterizing the horizon. However, 
this is made possible by the stress-energy tensor developing a singularity on 
the horizon \cite{marolfds}.

For a de Sitter boundary, 
we compute the entanglement entropy and explore its relation to the gravitational entropy. 
In order to obtain a better understanding of the properties of the thermalized CFT,
we also analyze a different minimal surface that corresponds to 
the world sheet of a string extending into the bulk from a single boundary point. 
Its area
allows us to extract the thermal contribution to the energy of a 
heavy quark located at the boundary point \cite{probequark} and read off the 
corresponding temperature. 
In order to establish the universality of our findings, 
we repeat the calculation of the entropy for a black-hole boundary metric with a mass parameter different from that of the bulk black hole \cite{marolfbh}.
Finally, we derive an expression that applies to any boundary metric with 
a horizon.

The plan of the paper is the following: 
In section \ref{BTZsection} we review the case of a bulk black hole for 
a flat AdS boundary. We determine the relevant minimal surfaces for the 
calculation of the entropy and the CFT temperature, 
and discuss the issue of the definition of Newton's constant. 
In section \ref{dssection}, through an appropriate choice of coordinates,  
we derive a metric with a dS boundary and a bulk black hole, dual to 
a thermal field theory in curved spacetime. 
We discuss analytically the entangling surface, the divergent terms 
and the dependence of the entropy on  the ratio of the thermal CFT temperature 
$T$ and the de Sitter temperature $T_{dS}$. 
We also determine analytically the minimal surface that determines 
the energy associated with a heavy quark on the boundary, in order to deduce
the temperature from the thermal contribution.
The qualitative behavior of all minimal surfaces turns out to be 
related to the ratio $T/T_{dS}$ that determines 
the relative strength of the gravitational potentials 
generated by the black-hole and cosmological horizons.
In section \ref{bhsection} we repeat the analysis 
for a boundary with a black-hole horizon. Finally, in 
section \ref{conclusions} we generalize our results 
for an arbitrary static boundary metric with horizons.

\section{Bulk black hole}
\label{BTZsection}

\subsection{Interpretation}

We consider solutions of the Einstein field equations in 2+1 dimensions with a
negative cosmological constant.  All such solutions are locally isometric to AdS space. We normalize all dimensionful parameters with respect to
the AdS length $l$, which is equivalent to setting $l=1$.

We start by considering a metric of the form
\begin{equation}
\label{eqmetric}
ds^2 = -(r^2-\mu) dt^2 + \frac{dr^2}{r^2-\mu} + r^2 d\phi^2~,
\end{equation}
with $r\in [\sqrt{\m},\infty]$. If $\mu$ vanishes,
eq. (\ref{eqmetric}) describes part of the covering space of AdS.
This becomes obvious if we observe that the definition of a new coordinate $u=1/r$ turns eq. (\ref{eqmetric}) into the standard  AdS metric in Poincare coordinates.
If $\mu$ is non-zero and
the coordinate $\phi$ is taken to be periodic, with period equal to $2\pi$,
the metric describes a nonrotating BTZ black hole \cite{btz} with mass parameter
 $\mu$ and
Hawking temperature $T = \sqrt{\mu}/(2\pi)$. %
In the following we allow $\phi$ to take values over the whole
real axis.
For the metrics we study, a $\mu$-dependent 
bulk horizon always exists, which allows us to 
make the standard assumption
that $\sqrt{\mu}$ is related to the temperature of the dual CFT.

An asymptotically AdS geometry can be related
to a dual CFT on the boundary through the AdS/CFT correspondence \cite{adscft1,adscft2}.
An efficient way to establish the connection is through the use of
Fefferman-Graham coordinates \cite{fg}.
The most general (2+1)-dimensional metric that satisfies Einstein's equations with
a negative cosmological constant is of the form \cite{skenderisbh}
\be\label{eq2} ds^2 = \frac{1}{z^2} \left[ dz^2 + g_{\mu\nu} dx^\mu dx^\nu \right]~, \ee
where
\be g_{\mu\nu} = g_{\mu\nu}^{(0)} + z^2 g_{\mu\nu}^{(2)}
+ z^4 g_{\mu\nu}^{(4)}~. \label{fefg} \ee
The stress-energy tensor of the dual CFT is \cite{skenderis}
\begin{equation}
\label{eq3a}
\langle T_{\mu\nu} \rangle =
\frac{1}{8\pi G_3}
\left[ g^{(2)}-{\rm tr}\left( g^{(2)}\right)g^{(0)} \right]~.
\end{equation}

The metric of eq. (\ref{eqmetric}) can be put in the form of eq. (\ref{eq2})
by defining a new coordinate
$z=(2/\mu)\left(r-\sqrt{r^2-\mu} \right)$.
It becomes
\begin{equation}
\label{eqmetric1}
ds^2 = \frac{1}{z^2}
\left[ dz^2 - \left( 1-\frac{\mu}{4}z^2\right)^2 dt^2 +\left( 1+\frac{\mu}{4}z^2\right)^2 d\phi^2 \right]~,
\end{equation}
where $z\in [0,z_h:=2/\sqrt{\m}]$ and the metric is of the desired form with
a flat boundary
\begin{equation}
\label{eqmetric0}
ds_0^2 = g_{\mu\nu}^{(0)} dx^\mu dx^\nu = -dt^2 +d\phi^2~.
\end{equation}
From eq. (\ref{eq3a}), we obtain the CFT energy density and pressure
\begin{eqnarray}
\label{eq3e}
\rho&=&\frac{E}{V}=-\langle T_{~t}^{t} \rangle=\frac{\mu}{16\pi G_3},
\\
p& =& \langle T_{~\phi}^{\phi} \rangle =
\frac{\mu}{16\pi G_3}.
\label{eq3p}
\end{eqnarray}

\subsection{Entanglement}
We are interested in computing the entanglement entropy of a part of the
boundary enclosed by an entangling surface $\Ac$.
According to the Ryu-Takayanagi conjecture \cite{ryu,review},
the entanglement entropy is proportional to the
area of a minimal surface
$\gamma_A$ extending from $\Ac$ into the bulk.
For the metric (\ref{eqmetric1}) the area to be minimized is actually a length,
given by
\be
{\rm Area}(\gamma_A)=\int_{\phi_1}^{\phi_2} d\phi \frac{1}{z} \sqrt{\left(\frac{dz}{d\phi}\right)^2 +\left( 1+\frac{\mu}{4}z^2\right)^2}.
\label{areabtzz} \ee
The minimization results in the differential equation
\be
z \left(1+\frac{1}{4}\mu\, z^2\right) z''+  \left(1-\frac{3}{4}\mu\, z^2\right) \left(z'\right)^2
 + \left(1-\frac{1}{4}\mu\, z^2\right) \left(1+\frac{1}{4}\mu\, z^2\right)^2=0.
\label{eqbtzmin} \ee
Its solution with boundary conditions $z(\phi_1)=z(\phi_2)=0$, with $\phi_2>\phi_1$, is
\be
z(\phi)=\frac{2}{\sqrt{\mu}}\sqrt{
\frac{\left( e^{\sqrt{\mu}\phi}-e^{\sqrt{\mu}\phi_1}\right)
\left( e^{\sqrt{\mu}\phi_2}-e^{\sqrt{\mu}\phi}\right)}
{\left( e^{\sqrt{\mu}\phi}+e^{\sqrt{\mu}\phi_1} \right)
\left( e^{\sqrt{\mu}\phi}+e^{\sqrt{\mu}\phi_2} \right)}
}.
\label{solbtzmin}
\ee
The solution (\ref{solbtzmin}) can also be expressed in coordinates $(r,\phi)$.
It reads
\be
r(\phi)=\frac{\sqrt{\mu}\, e^{\sqrt{\mu}\phi}
\left( e^{\sqrt{\mu}\phi_1}+e^{\sqrt{\mu}\phi_2}\right)
}{\sqrt{
\left( e^{2\sqrt{\mu}\phi}-e^{2\sqrt{\mu}\phi_1}\right)
\left( e^{2\sqrt{\mu}\phi_2}-e^{2\sqrt{\mu}\phi}\right)}}.
\label{solbtzminr} \ee

Substituting eq. (\ref{solbtzmin}) in eq. (\ref{areabtzz}) leads to a divergent
integral. There are two identical divergences, arising from the points
$\phi=\phi_1$ and  $\phi=\phi_2$ at
which the minimal curve approaches the boundary.
The integral can be made finite by imposing a cutoff at $z=\epsilon$.
The leading contribution is ${\rm Area}(\gamma_A)=2\log(1/\epsilon)$.
The entanglement entropy is given by \cite{ryu,review}
\be
S=\frac{{\rm Area}(\gamma_A)}{4G_{3}}=\frac{\log(1/\epsilon)}{2G_3},
\label{entropybh} \ee
displaying the characteristic dependence on the UV cutoff.

\subsection{Effective Newton's constant}

Expressions such as eq. (\ref{entropybh}) 
for the entanglement entropy arise also in cases in which
the boundary metric is non-trivial. We are especially interested in situations
in which this metric has a horizon. If the entangling surface coincides
with the horizon, one would expect the entropy to be related to the
gravitational entropy.
In order to assign a physical meaning to a divergent expression, such as
eq. (\ref{entropybh}),
one must define an effective
Newton's constant for the boundary theory. Such a definition was given
in refs. \cite{giataganas,correctionsds}, in the spirit of ref. \cite{hawking}.
The effective Newton's constant for a $(d+1)$-dimensional boundary theory is
\be
G_{d+1}=(d-1) \epsilon^{d-1} G_{d+2},
\label{Geff} \ee
with $(d-1)\epsilon^{d-1}$ replaced by 1/$\log(1/ \epsilon$) for $d=1$.
In four dimensions this definition can be justified in the context of
the Randall-Sundrum (RS) model \cite{rs}. Notice that in the limit
$\epsilon\to 0$ the constant vanishes and gravity becomes non-dynamical.
This makes it difficult to compute the gravitational entropy in
the context of the AdS/CFT correspondence.

An alternative, more rigorous way to obtain the same result is through
holographic renormalization
\cite{skenderis,papadimitriou}, which produces the
stress-energy tensor of the dual CFT discussed above.
The bulk metric of a five-dimensional
asymptotically AdS space is written in a
Fefferman-Graham expansion \cite{fg} in terms of the bulk coordinate $z$.
A solution is then obtained order by
order. The on-shell gravitational action is
regulated by restricting the bulk integral to the region $z>\epsilon$.
The divergent terms are subtracted through the introduction of
appropriate counterterms. In this way a renormalized effective action
is obtained, expressed in terms of the induced metric
$\gamma_{ij}$ on the surface at $z=\epsilon$.
In our approach the entropy is not
renormalized. We assume that the cutoff $\epsilon$ is physical and we
incorporate it in the effective couplings.
This amounts to employing the regulated form of the effective action,
without the subtraction of divergences.
The leading terms, which would diverge for $\epsilon \to 0$,
can be found in the counterterm action of
holographic renormalization. They are expressed in terms of the induced metric
$\gamma_{ij}$,
which includes a factor
$\epsilon^{-2}$.
Using the results of \cite{skenderis, papadimitriou} and
extracting the $\epsilon^{-2}$ factor
from $\gamma_{ij}$, we can express the leading terms of
the regulated action as
\be
S=\frac{1}{16\pi G_3}\int d^2x\, \sqrt{-\gamma}
\left[ \frac{2}{\epsilon^2}-\log(\epsilon) {\cal R} \right].
\label{effaction} \ee
The first term corresponds to a cosmological constant, which must be
(partially) cancelled by vacuum energy localized on the surface at $z=\epsilon$,
such as the brane tension in the RS model \cite{rs}.
The second term is the standard Einstein term if the effective Newton's
constant $G_2$ is defined as
\be
G_2=\frac{G_3}{ \log (1/\epsilon)}.
\label{G2def} \ee


\subsection{Thermal effects} \label{thermaleffects}

In this section we examine 
the thermal effects of the strongly-coupled fields sourced 
by a heavy probe quark located on the boundary. The total 
energy of the quark can be obtained from the world-sheet area of a string
attached to the quark and extending into the bulk \cite{probequark}.
One expects an infinite contribution from the region near the boundary, which 
needs to be regularized, and a contribution from the other endpoint that is attached to 
the bulk horizon, which carries the information on the temperature. 
The string solution with world-sheet parameters $\prt{\t,\s}$ 
corresponding to a quark in the coordinate system \eq{eqmetric} 
is given by a radial-gauge parametrization of the form $\prt{t,r}=\prt{\t,\s}$, 
where the string hangs from a fixed point with $\phi=\phi_b$
on the boundary at $r\rightarrow \infty$. For a flat
boundary, the string remains straight as it extends 
into the bulk. 
The surface can be parametrized in a similar fashion in the 
coordinate system \eq{eqmetric1}. 
In this case we have $\prt{t,z}=\prt{\t,\s}$, 
where the string now hangs from a point 
$\phi=\phi_b$ on the boundary at $z\rightarrow 0$.

The minimal surface is a solution of the  Nambu-Goto equations of motion. 
Its area includes a multiplicative time factor $\cT$ and is given by 
\be\la{m1}
S=\ff{\cT}{2 \pi \a'}\int_{z_b}^{z_h} dz \sqrt{-g_{tt} g_{zz}}=\ff{\cT}{2 \pi \a'}\int_{z_b}^{z_h}  \ff{dz}{4z^2}\prt{4-\m z^2}~,
\ee
where we impose a cutoff $z_b=\epsilon$, while $z_h=2/\sqrt{\m}$ is the deepest point that the string can reach in the bulk because of the presence of the 
black-hole horizon. 
The (rescaled) energy $E$ of the string, arising from the strongly-coupled 
fields sourced by the quark, is given by
\be\la{m2}
E:=\frac{2 \pi \a'}{\cT} S=\frac{1}{\epsilon}- \sqrt{\m} +{\cal O}(\epsilon)~.
\ee
The thermal effects result in a leading contribution proportional 
 to $\sqrt{\mu}$ and, therefore, to the temperature.

\section{Static de Sitter boundary}
\label{dssection}

In this section we study a thermal field theory of temperature $T$ in a 
de Sitter spacetime with a different temperature $T_{dS}$ associated with 
the cosmological horizon. 
The dual picture involves a metric with  
a de Sitter boundary in static coordinates in the 
presence of black-hole configuration in the bulk.
The metric (\ref{eqmetric}) can be put in the form
\begin{eqnarray}
ds^2
= \frac{1}{z^2} \Bigg[ dz^2
&-& (1-H^2\rho^2)\left(1-\frac{1}{4}\left[\frac{\mu-H^2}{1-H^2\rho^2}+H^2\right]z^2 \right)^2 dt^2
\Biggr.
\nonumber \\
\Biggl.
&+& \left(1+\frac{1}{4}\left[\frac{\mu-H^2}{1-H^2\rho^2}-H^2\right]z^2 \right)^2  \frac{d\rho^2}{1-H^2\rho^2} \Biggr]~,
\label{eqmetric5} \end{eqnarray}
with a de Sitter boundary metric
\begin{equation}
\label{eqmetric300}
ds_0^2 = g_{\mu\nu}^{(0)} dx^\mu dx^\nu = -(1-H^2 \rho^2) dt^2 + \frac{d\rho^2}{1-H^2\rho^2}~.
\end{equation}
We concentrate on the causally connected region $-1/H < \rho < 1/H$.
The space  contains two cosmological horizons (and their extensions into the bulk)
at $\r_h=\pm 1/H$, and a bulk horizon at
\be\la{ds_hor}
z_h(\r)=\frac{2\sqrt{1-H^2\rho^2}}{\sqrt{\m-H^4 \r^2}}~.
\ee
It is apparent that the bulk horizon extends smoothly to the location of
the cosmological horizon on the boundary only for $\mu \geq H^2$.
For this reason, we consider only this parameter range in the following. 
The deepest point of the bulk horizon is at $z_{h, max}=2/\sqrt{\m}$  for $\rho=0$. 
For $\mu>H^2$, the cosmological horizons are located behind the bulk horizon.

The metric of eq.
(\ref{eqmetric5}) can be derived from that of eq. (\ref{eqmetric}) with a coordinate transformation that does not affect the time coordinate. The
transformation is
\begin{eqnarray}
r(z,\rho)&=&\sqrt{\mu+\frac{\left[ 1-H^2\rho^2-(\mu-H^4\rho^2)z^2/4\right]^2}{z^2(1-H^2\rho^2)} }~,
\label{bb22} \\
\phi(z,\rho)&=&\frac{1}{2H}\log\left[\frac{1+H\rho}{1-H\rho} \right]
+\frac{1}{2\sqrt{\mu}}\log\left[\frac{1-H^2\rho^2+\left(\sqrt{\mu}-H^2\rho \right)^2z^2/4}{1-H^2\rho^2+\left(\sqrt{\mu}+H^2\rho \right)^2z^2/4} \right].
\label{bb3aa}
\end{eqnarray}
The region near negative infinity for $\phi$ is mapped
to the vicinity of $-1/H$ for $\rho<0$, and the region near positive infinity for
$\phi$ to the vicinity of $1/H$ for $\rho>0$. It is apparent that $\phi$ cannot be
considered to be periodic.

The stress-energy tensor corresponding to the metric (\ref{eqmetric5}) is equal to
\begin{eqnarray}
\label{eq3eeiii}
\rho&=&-\langle T_{~t}^{t} \rangle=
\frac{1}{16\pi G_3}\left(\frac{\mu-H^2}{1-H^2\rho^2}-H^2 \right)
\\
p& =& \langle T_{~\rho}^{\rho} \rangle =
\frac{1}{16\pi G_3}\left(\frac{\mu-H^2}{1-H^2\rho^2}+H^2 \right)~,
\label{eq3epiii}
\end{eqnarray}
where the presence of the conformal anomaly is evident, as
\begin{equation}
 \langle T_{~\mu}^{\mu} \rangle  =
\frac{1}{8\pi G_3}H^2.
\label{confaneiii}\end{equation}
For $\mu > H^2$ the stress-energy tensor displays singularities at the locations of the horizons $\rho=\pm1/H$ of static de Sitter space. As also discussed in ref. \cite{marolfds}, we interpret this configuration as a boundary CFT
at a temperature $T=\sqrt{\mu}/(2\pi)$
larger than the de Sitter temperature $T_{dS}=H/(2\pi)$. 
The divergence of the energy density on the de Sitter horizon can be viewed as a
heat bath localized at this point that keeps the system at a 
temperature different from the de Sitter temperature.

\subsection{Entropy} \label{entropyds}
\begin{figure}[t]
\centerline{\includegraphics[width=0.7\textwidth]{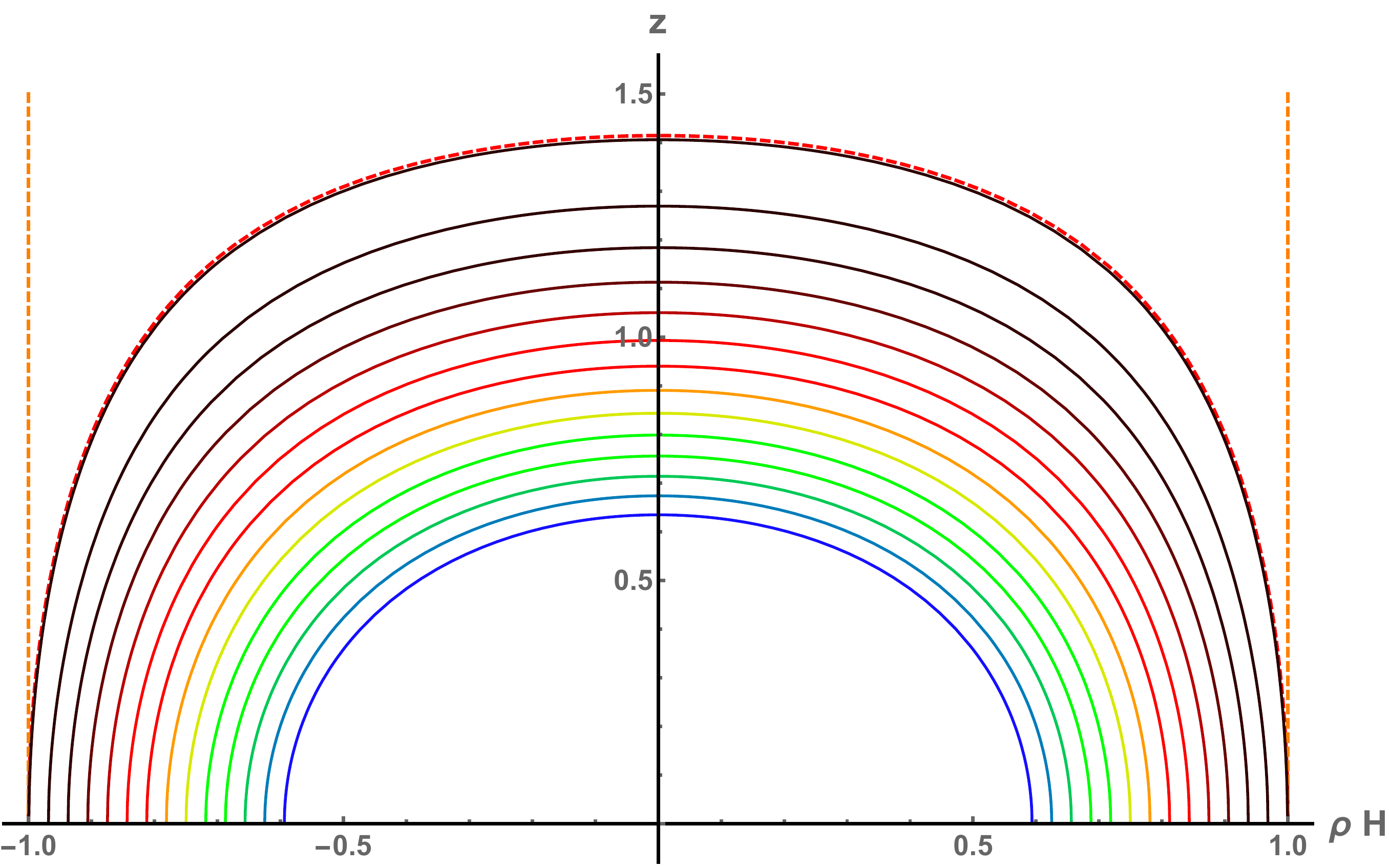}}
\caption{\small The minimal curves for $\mu=2H^2$ and for 
different entangling boundary 
points. The bulk horizon of eq. \eq{ds_hor} is denoted by the curved dashed line,
and the cosmological horizons at $\rho=\pm 1/H$ by the vertical dashed lines. 
When the endpoints approach the cosmological horizons, 
the entangling surface coincides with 
the bulk horizon.   \label{fig:1}}
\end{figure}

We turn next to the discussion of the entropy associated with such a configuration.
In the holographic approach the entanglement entropy is given by a spacelike
minimal curve anchored on the two points on the boundary that act as the
entangling surface. The calculation can be performed by minimizing the
functional
\be
{\rm Area}(\gamma_A)=
\int_{-1/H}^{1/H} d\rho \frac{1}{z} \sqrt{ \left( \frac{dz}{d\rho}\right)^2
+ \left(1+\frac{1}{4}\left[\frac{\mu-H^2}{1-H^2\rho^2}-H^2\right]z^2 \right)^2  \frac{1}{1-H^2\rho^2} },
\label{areadsbh} \ee
Finding an analytical solution of the resulting 
non-linear second order equation is a difficult problem. 
However, it becomes tractable if one uses the fact that
the minimal curve is a geometrical object which does not depend on the system of
coordinates. This means that one can simply use the solution of eq. (\ref{solbtzminr})
and transform the coordinates through eqs. (\ref{bb22}), (\ref{bb3aa}).
The resulting implicit relation between the coordinates $z$ and $\rho$ can be
solved explicitly in this case. By imposing the appropriate 
boundary conditions we obtain
\bea\nn
&&z(\rho)=2 \Bigg[\bigg[\left(H^2 \rho^2-1\right) \left(H^2 \rho +\sqrt{\mu }\right)^2 F(\rho_1) F(\rho_2)
\\ \nn
&&+2 \sqrt{\mu}  \left(1-H^2 \rho ^2\right) F(\rho)^\frac{1}{2} \left(F(\rho_1 )^{\frac{1}{2}}+F(\rho_2 )^{\frac{1}{2}}\right) \left(\left(H^2 \rho +\sqrt{\mu }\right) F(\rho_1)^{\frac{1}{2}} F(\rho_2)^{\frac{1}{2 }}+\left(\sqrt{\mu }-H^2 \rho \right) F(\rho )\right)
\\ \nn
&&+\left(H^2 \rho ^2-1\right) \left(\sqrt{\mu }-H^2 \rho \right)^2 F(\rho )^2\\\nn
&&-\left(H^2 \rho ^2-1\right) F(\rho ) \left(\left(H^4 \rho ^2-\mu \right) F(\rho_1)+\left(H^4 \rho ^2-\mu \right) F(\rho_2)-4 \mu  F(\rho_1)^{\frac{1}{2}} F(\rho_2)^{\frac{1}{2 }}\right)\bigg]/\\\nn
&&\bigg[\left(\left(\sqrt{\mu }-H^2 \rho \right)^2 F(\rho )-\left(H^2 \rho +\sqrt{\mu }\right)^2 F(\rho_1)\right) \left(\left(\sqrt{\mu }-H^2 \rho \right)^2 F(\rho )-\left(H^2 \rho +\sqrt{\mu }\right)^2 F(\rho_2)\right)\bigg]\Bigg]^\frac{1}{2}.
\label{z1}
\eea
with
\be
F(\rho):=\left( \frac{1-H \rho}{1+H\rho} \right)^\frac{\sqrt{\mu}}{H}~.
\label{frho}\ee
The minimal curve approaches the boundary $z=0$ at $\rho=\rho_1$ and
$\rho=\rho_2$. In Figure \ref{fig:1} we present minimal curves
for a range of entangling boundary segments, with lengths that increase until 
the whole region between the boundary horizons is covered. In this limit
 the minimal curve coincides with the bulk horizon.

The length of the minimal curve receives an infinite contribution from the
region near the boundary and must be regulated through the introduction of an
appropriate cutoff at $z=\epsilon$. 
It must be pointed out that the only element that differentiates the final result for
the various metrics that we are considering is the way the divergence near the boundary is regulated.
Our basic assumption is that the cutoff must be imposed on the
bulk coordinate in the Fefferman-Graham parametrization. Even though this
coordinate is always denoted by $z$, setting $z=\epsilon$ implements a
different cutoff procedure for every choice of boundary metric. This is
reflected in quantities that depend on the cutoff, such as the length of
the minimal curve that determines the entanglement entropy.

We are interested in the limit that the entangling surface covers the
whole region between the horizons of the boundary (1+1)-dimensional
de Sitter metric. These are located at $\rho_1=-1/H$ and $\rho_2=1/H$.
The resulting entropy results from the entanglement between the
``North" and ``South" static patches of the global geometry (\cite{giataganas,correctionsds}).
In this limit, the minimal curve (\ref{z1}) coincides with the bulk
horizon of the metric of eq. (\ref{eqmetric5}), which
is given by eq. \eq{ds_hor}
The length of the minimal curve can be computed by substituting this expression
in eq. (\ref{areadsbh}). We focus on the two divergent contributions, arising from
the vicinity of the boundary. Imposing a cutoff at $z=\epsilon$, we find
\be
{\rm Area}(\gamma_A)=2
\frac{\sqrt{\mu}}{H}\log\left( \frac{1}{\epsilon} \right)
+{\rm finite~terms}~.
\label{areadsbhcutoff} \ee
The entanglement entropy is
\be
S_{\rm dS}=\frac{{\rm Area}(\gamma_A)}{4G_{3}}=
\frac{\sqrt{\mu}}{H}\frac{\log(1/\epsilon)}{2G_3}~.
\label{entends} \ee

Making use of eq. (\ref{G2def}) allows us to write eq. (\ref{entends}) as
\be
S_{\rm dS}=
\frac{\sqrt{\mu}}{H}\frac{2}{4G_2}~.
\label{entendss} \ee
For $\sqrt{\mu}=H$ this is the expected gravitational entropy of the
(1+1)-dimensional de Sitter space, whose horizons are two points
\cite{hawking}. The bulk black hole corresponds to a dual CFT at the same
temperature as the horizons. For $\sqrt{\mu}>H$, the above expression
seems to indicate a modification of Bekenstein's relation between the
entropy and the area of the horizon \cite{bekenstein}. This can be
attributed to the divergence of the stress-energy tensor at the horizon,
with a singular (positive) energy density given by eq. (\ref{eq3eeiii}).
However, the situation can also be viewed differently.
In ref. \cite{correctionsds} it was
shown, using holography, that the de Sitter entropy in 3+1 dimensions
receives subleading logarithmic corrections, regulated by the
UV cutoff, which are proportional to the central
charge of the dual CFT. These corrections arise through the higher curvature terms
in the
effective action responsible for the conformal anomaly. In
1+1 dimensions we expect similar logarithmic corrections. However, these
are now of the same order as the leading contribution to the gravitational
entropy. In this spirit, eq. (\ref{entends}) could be rewritten as
\be
S_{\rm dS}=\frac{2}{4G_2}+ \frac{1}{2G_3}
\left(\frac{\sqrt{\mu}}{H}-1\right)\log\left( \frac{1}{\epsilon} \right)~,
\label{entendsss} \ee
where the coefficient of the logarithm accounts for degrees of freedom
of a CFT thermalized above the horizon temperature.

\subsection{Thermal effects} \label{thermaleffects2}

In order to analyze the temperature of the CFT from a different perspective, we 
consider the thermal corrections to the energy associated with a heavy probe
quark at some point on the boundary of a bulk space described by the metric \eq{eqmetric5}. 
As we reviewed in subsection \ref{thermaleffects}, 
the energy can be obtained through the minimization of the Nambu-Goto action 
of a string with a single boundary endpoint at the location of the proble quark \cite{probequark}.
When the quark is placed at $\r_b=0$, the string extends along a straight line from
the boundary to the bulk horizon. The energy is given by eq. \eq{m2}. 
The thermal effects are accounted for by a temperature $T=\sqrt{\m}/(2\pi)$.
As we assume that $\sqrt{\mu}\geq H$, 
this temperature satisfies $T\geq T_{dS}$, with the two coinciding only for 
$\sqrt{\m}=H$.

If the string endpoint is placed at $\r_b\neq0$, 
a straight line configuration
does not minimize the Nambu-Goto action any more. 
The gravitational influence of the
horizons is now uneven, and the string  
bends toward the closest cosmological horizon. 
The equations for the minimization problem in this case are obtained from 
the action
\bea\la{min_btzds}
&&S=\frac{\cT}{2\pi \a'} \int d\rho \frac{1}{z^2}\sqrt{h(\r) G_{-}^2(\r,z) \prt{z'(\rho)^2+ \frac{G_{+}^2(\r,z)}{h(\rho)}}}~,\\\nn
&&h(\r):=1-H^2\rho^2~,\qquad G_{\pm}(\r,z):=1\mp\ff{1}{4}\prt{\frac{\m-H^2}{h(\r)}\pm H^2}z^2~,
\eea
and are lengthy non-linear differential equations of second order.
We take advantage of the fact that we know the straight string 
solution for the metric \eq{eqmetric}, 
which is given by the parametrization $\prt{t,r,\phi}=\prt{\t,\s,\phi_b}$, 
with  constant $\phi_b$. 
From this we can obtain the corresponding solution 
for the metric (\ref{eqmetric5})
by using eqs. \eq{bb22} and \eq{bb3aa}. 
The  surface that satisfies the equations of motion arising
from minimizing the functional \eq{min_btzds} reads
\be\la{wl_sol}
z(\rho)=\frac{2\sqrt{h(\r)}\sqrt{F(\r)-F(\r_b)}}{\sqrt{-F(\r)\prt{\sqrt{\m}+H^2 \r}^2+\prt{\sqrt{\m}-H^2 \r}^2 F(\r_b)}}~,
\ee
where the function $F(\rho)$ has been defined in eq. (\ref{frho}).
In our coordinate system, $\r_b$ is the boundary position of the string, $\r_b\in[0,1/H]$, which implies that $F(\r_b)\in[0,1]$.
The coordinates $\r_b$ and $\phi_b$ are related through 
$F(\r_b)=e^{-2 \sqrt{\m} \phi_b}$. The point $\rho_h$ at which the
minimal surface hits the bulk horizon 
can be found by solving the equation $z(\rho_h)=z_h(\rho_h)$, with the horizon $z_h(\r)$ given  by eq. \eq{ds_hor}.

\begin{figure}[t]
\centerline{\includegraphics[width=0.52\textwidth]{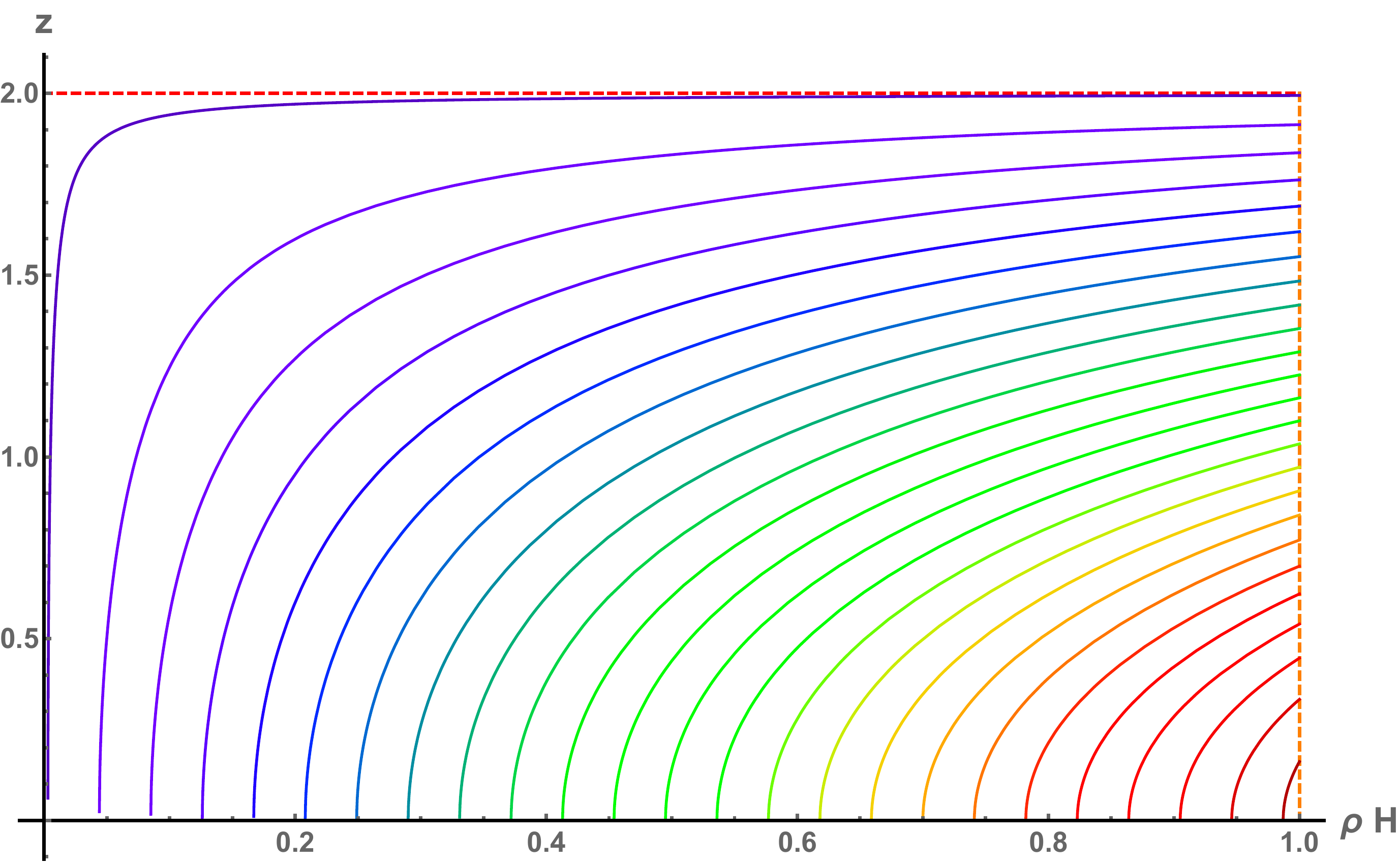}~\includegraphics[width=0.51\textwidth]{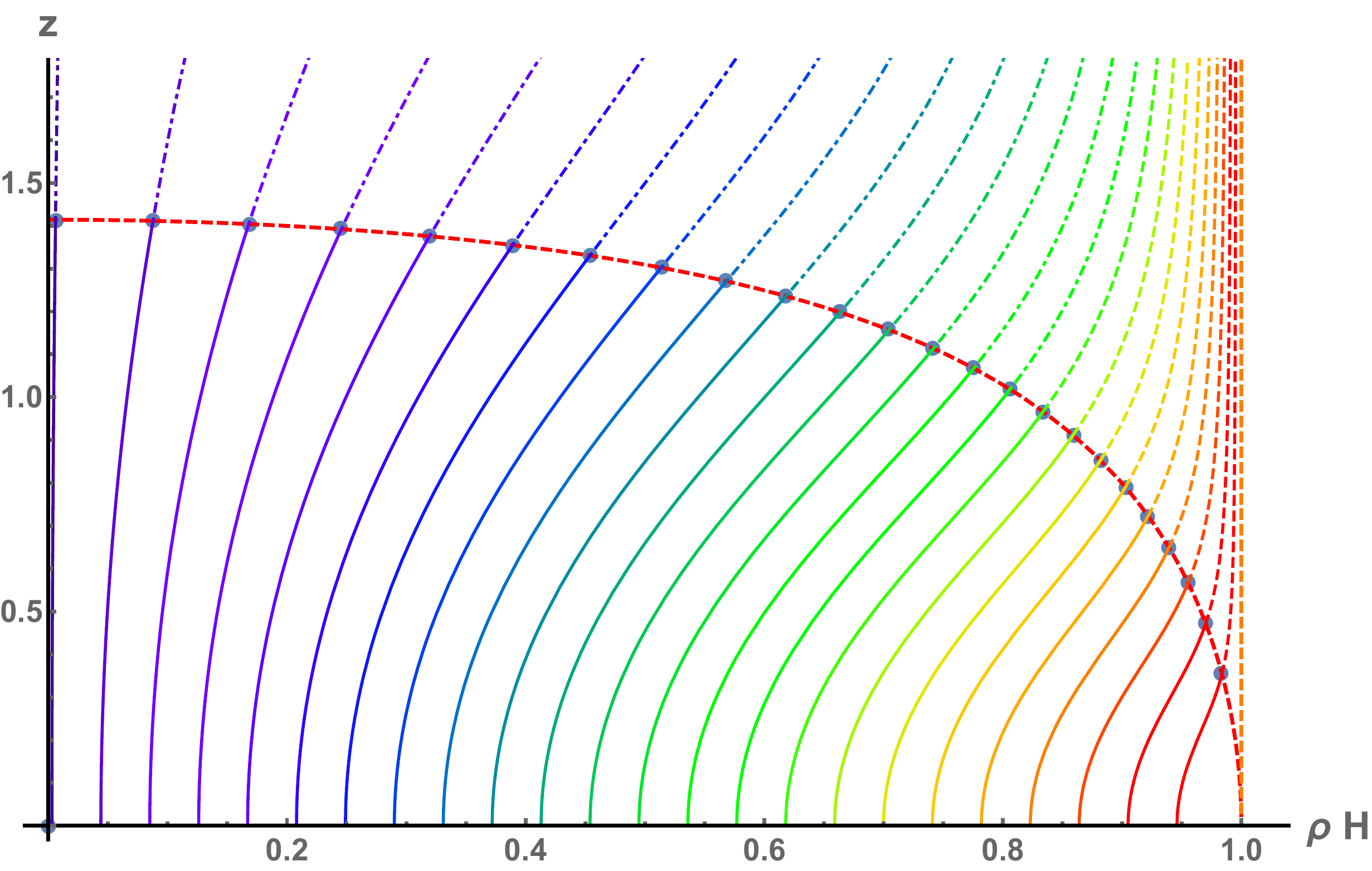}}
\caption{\small The minimal surfaces for several boundary points between $\r_b=0$
and the cosmological horizon at $\r_b= H^{-1}$. \textbf{Left plot}: $\m=H^2$. 
The 
bulk horizon at $z=2/\sqrt{\m}$ is depicted as a horizontal dashed line and the cosmological 
horizon at $\rho=H^{-1}$ as a vertical dashed line. 
\textbf{Right plot}: $\m\neq H^2$.  
The bulk horizon, given by eq. \eq{ds_hor}, is depicted as a curved 
dashed line. 
The dot-dashed extensions of the minimal surfaces beyond the bulk horizon 
are not relevant for our purposes.  \label{fig:2}}
\end{figure}

We consider first the special case $\m=H^2$, for which 
the CFT temperature is equal to the de Sitter temperature \cite{Chu:2016pea}. 
The minimal surface simplifies to
\be\la{zeq1}
z(\rho)=\frac{2}{H}\sqrt{\frac{ \r-\r_b} {\r+\r_b}},
\ee
starting from the boundary point $\rho_b$ and hitting the cosmological horizon at
\be\la{zeh1}
z(H^{-1})=\ff{2}{H} \sqrt{F(\r_b)}~.
\ee
The form of the surface is depicted in the left plot of Figure \ref{fig:2}.  
Surfaces that start from $\r_b\simeq0$ tend to approach quickly the bulk 
horizon before turning towards the cosmological one. 
On the other hand, surfaces with $\rho_b\simeq H^{-1}$ 
bend towards the cosmological horizon without going deep into the bulk.
The energy can now be computed by performing the integration 
in eq. \eq{min_btzds} and cutting off the divergent contribution from the 
lower limit at $z=\epsilon$. We find
\be
E=-\frac{\sqrt{1-H^2 \r_b^2}}{z(\r)}\prt{1+\frac{1}{4}H^2 z(\r)^2}\bigg|_{\r_b}^{H^{-1}}
=\frac{\sqrt{1-H^2 \r_b^2}}{\e} -H +{\cal O}(\epsilon)~.
\ee
The divergent part receives a correction corresponding to 
the redshift factor of the boundary metric. 
The finite term is proportional to the temperature $T_{dS}$, as expected.

We turn now to the general case $\mu>H^2$. 
The string described by eq. \eq{wl_sol} hits the bulk horizon 
at a point $\r_h$ satisfying the algebraic equation $z(\r_h)= z_h(\rho_h)$,
where  $z_h(\r)$ is given by eq. \eq{ds_hor}. 
An implicit solution is given by the relation
\be\la{csol1}
F(\rho_h)\frac{\sqrt{\m}+H^2 \r_h }{\sqrt{\m}-H^2 \r_h}=F(\rho_b)~,
\ee
which can be solved explicitly for $\rho_b$.
The leading terms of the above expression  near the origin ($\r_b\simeq 0$), 
and near the cosmological horizon ($\r_b\simeq H^{-1}$), respectively read
\bea
\r_h{}_{(0)}=\frac{\m}{\m-H^2}\r_b ~,
\qquad \r_h{}_{(H^{-1})}= H^{-1}+\prt{\frac{\sqrt{\m}+ H}{\sqrt{\m}-H}}^{-\frac{H}{\sqrt{\m}}}\prt{\rho_b-H^{-1}}~.
\eea
The integration of eq. \eq{min_btzds} can be performed analytically 
and, after some algebraic on-shell manipulations, gives
\be\la{m_btzds}
E=- 2 H^2 \r  \sqrt{
\frac{ \m\, F(\r)F(\r_b)}{\prt{F(\r)-F(\r_b)}\prt{-F(\r)\prt{\sqrt{\m}+H^2 \r}^2+\prt{\sqrt{\m}-H^2 \r}^2F(\r_b)}}}\Bigg|_{\r_b}^{\r_h}~.
\ee
The use of eq. \eq{csol1} demonstrates that the upper limit gives a 
finite contribution equal to $-\sqrt{\mu}$. The lower limit gives a
divergent contribution, which is regulated by imposing a cutoff at $z=\epsilon$,
using an expansion of the solution (\ref{wl_sol}) around $\rho_b$.
In this way we find 
\be\la{Ewl_sol}
 E=\frac{\sqrt{1-H^2\r_b^2}}{\e}-\sqrt{\m}+{\cal O(\epsilon)}. 
\ee
The divergent term contains the same redshift factor found earlier, while 
the finite contribution arises through $\sqrt{\mu}.$ This result confirms that
the relevant scale for the CFT temperature is set by the mass of the
bulk black hole.

\section{Black-hole boundary}
\label{bhsection}

It is possible to define appropriate coordinates such that
the metric on the boundary becomes that of a (1+1)-dimensional black hole.
The form of the metric for an asymptotically AdS (2+1)-dimensional
spacetime with an arbitrary static boundary metric
has been derived in ref. \cite{skenderisbh}. Exploiting this result,
we consider a metric of the form
\begin{eqnarray}
ds^2
= \frac{1}{z^2} \Bigg[ dz^2
&-& (1-e^{-2x})\left(1-\frac{\mu-e^{-4x}}{4(1-e^{-2x})}z^2 \right)^2  dt^2
\Biggr.
\nonumber \\
\Biggl.
&+&\left(1-e^{-2x}z^2+\frac{\mu-e^{-4x}}{4(1-e^{-2x})}z^2 \right)^2 \frac{dx^2}{1-e^{-2x}} \Biggr]~,
\label{eqmetricbh} \end{eqnarray}
with a black-hole boundary metric
\begin{equation}
\label{eqmetricbhb}
ds_0^2 = g_{\mu\nu}^{(0)} dx^\mu dx^\nu = -(1-e^{-2x}) dt^2 + \frac{dx^2}{1-e^{-2x}}
\end{equation}
that has a horizon at $x=0$ and a mass parameter equal to 1.

Through the coordinate change
\be
\rt={\rm arccosh}(e^x)~,
\label{transfrx} \ee
the metric can be written as
\be
ds^2
= \frac{1}{z^2}
\left( dz^2 - f(\rt,z)\, dt^2 +g(\rt,z)\,d\rt^2] \right),
\label{eqmetricbhr} \ee
where
\begin{eqnarray}
f(\rt,z)&=&\tanh^2(\rt)\left(1-z^2 \frac{1}{4\tanh^2(\rt)}\left(\mu-\frac{1}{\cosh^4(\rt)} \right)  \right)^2~,
\label{frz} \\
g(\rt,z)&=& \left(1+z^2 \frac{1}{4\tanh^2(\rt)}\left(\mu-\frac{4}{\cosh^2(\rt)}+\frac{3}{\cosh^4(\rt)} \right) \right)^2~.
\label{grz} \end{eqnarray}
For $\mu=1$ this metric reduces to the one
discussed in ref. \cite{marolfbh}.
It is apparent that covering the full range of positive and negative values of
$\rt$ requires two copies of the metric (\ref{eqmetricbh}) with $x$ taking
positive values.

The holographic stress-energy tensor for this metric is
\begin{eqnarray}
\label{eq3eeii}
\rho&=&-\langle T_{~t}^{t} \rangle=
\frac{1}{16\pi G_3}
\frac{4-3\, e^{-2x}-\mu \,e^{2x}}{1-e^{2x}}~,
\\
p& =& \langle T_{~x}^{x} \rangle =
\frac{1}{16\pi G_3}
\frac{e^{-2x}-\mu\, e^{2x}}{1-e^{2x}}~, 
\label{eq3epii}
\end{eqnarray}
displaying singularities at the location of the
black-hole horizon at $x=0$.
An expansion around this point gives
\begin{eqnarray}
\rho&=&\frac{1}{32\pi G_3}\left( \frac{\mu-1}{x}+\mu-5 \right)+{\cal O}(x)~,
\\ \label{emhor1}
p&=&\frac{1}{32\pi G_3}\left( \frac{\mu-1}{x}+\mu+3 \right)+{\cal O}(x)~.
\label{emhor2} \end{eqnarray}
The conformal anomaly is
\begin{equation}
 \langle T_{~\mu}^{\mu} \rangle  =
\frac{e^{-2x}}{4\pi G_3}~.
\label{confaneii}\end{equation}

The transformations that connect the coordinates $(z,x)$ with the coordinates
$(r,\phi)$ of eq. (\ref{eqmetric}) are
\begin{eqnarray}
r(z,x)&=&\sqrt{\mu+\frac{\left[ 1-e^{-2x}-(\mu-e^{-4x})z^2/4\right]^2}{z^2(1-e^{-2x})} }~,
\label{bb22bh} \\
\phi(z,\rho)&=&\frac{1}{2}\log\left[-1+e^{2x} \right]
-\frac{1}{2\sqrt{\mu}}
\log\left[\frac{z^2-2e^{2x}(2+\sqrt{\mu}\,z^2)+e^{4x}(4+\mu\, z^2)}{z^2+2e^{2x}(-2+\sqrt{\mu}\,z^2)+e^{4x}(4+\mu\, z^2)}
 \right],
\label{bb3aabh}
\end{eqnarray}
with the time coordinate remaining unaffected.

We are interested in an entangling surface,
which covers the whole region $x>0$  on the constant-time slice $t=0$.
The interior of this region is entangled with
a symmetric region in the global geometry, again parametrized with $x>0$
(but negative $\rt$). Both regions are located outside
the future and past horizons. Similarly to the case of a de Sitter boundary
metric, the minimal surface, which is a curve for a (1+1)-dimensional boundary,
starts on the boundary ($z=0$) at the location of the
horizon $x=0$. However, as there is no other horizon on the boundary,
the minimal surface must extend into the bulk indefinitely.
As a result, it coincides with the bulk horizon
\be
z(\rho)=2\sqrt{\frac{1-e^{-2x}}{\mu-e^{-4x}}}~.
\label{bulkhorbh} \ee
In the context of the Ryu-Takayanagi proposal, the entanglement entropy is
properly defined if the minimal curve returns to the boundary at a value
$x_r\to \infty$. However, the contribution to the length from the return point
is not relevant for our purposes, because it accounts for the entanglement with
the region $x>x_r$ for any finite value of $x_r$. This has been already
discussed in the context of a Rindler boundary in ref. \cite{giataganas}.
The upshot of these considerations is that the entanglement between the
two regions of the $t=0$ slice of the global geometry around the black-hole
horizon is accounted for by the length of the bulk horizon in the vicinity of
$x=0$.
By substituting eq. (\ref{bulkhorbh}) in the expression
\be
{\rm Area}(\gamma_A)=
\int_{0} dx \frac{1}{z} \sqrt{ \left( \frac{dz}{dx}\right)^2
+ \left(1-e^{-2x}z^2+\frac{\mu-e^{-4x}}{4(1-e^{-2x})}z^2 \right)^2 \frac{1}{1-e^{-2x}}},
\label{areabhbh} \ee
and keeping only the contribution from the lower limit, with a cutoff at $z=\epsilon$,
we find
\be
{\rm Area}(\gamma_A)=
\sqrt{\mu}\log\left( \frac{1}{\epsilon} \right).
\label{areabhbhcutoff} \ee
The entanglement entropy is
\be
S_{\rm bh}=\frac{{\rm Area}(\gamma_A)}{4G_{3}}=
\sqrt{\mu}\frac{\log(1/\epsilon)}{4G_3}.
\label{entenbh} \ee

Similarly to the de Sitter entropy, the black-hole entropy can be written,
through use of eq. (\ref{G2def}),
in two equivalent ways:
\be
S_{\rm bh}=\sqrt{\mu}\frac{1}{4G_2}=\frac{1}{4G_2}+ \frac{1}{4G_3}
\left(\sqrt{\mu}-1\right)\log\left( \frac{1}{\epsilon} \right),
\label{entendsbhh} \ee
with the same physical interpretation as the one discussed at the end
of subsection \ref{entropyds}. The analysis of the thermal contributions to the energy
of a probe quark leads to the conclusion that the relevant temperature scale is
set by the mass term $\sqrt{\mu}$ of the bulk black hole, in agreement with the findings
of subsection \ref{thermaleffects2}.

\section{Discussion}
\la{conclusions}

In the two previous sections we discussed specific examples of
boundary metrics with horizons, which made it possible to derive explicit
expressions for the minimal surfaces and the associated holographic
entropy. It would be very helpful to have a more global understanding
of the resulting expressions for a general boundary metric.
The most general metric for a (2+1)-dimensional AdS bulk containing
a non-rotating black hole and having a static boundary can be written
in the form \cite{skenderisbh}
\begin{eqnarray}
ds^2
= \frac{1}{z^2} \Bigg[ dz^2
&-& h(x)\left(1+\frac{1}{16}\frac{h'^2(x)-4\mu}{h(x)}z^2\right)^2  dt^2
\Biggr.
\nonumber \\
\Biggl.
&+&\left(1+\frac{1}{4}h''(x)z^2-\frac{1}{16}\frac{h'^2(x)-4\mu}{h(x)}z^2\right)^2 \frac{dx^2}{h(x)} \Biggr],
\label{eqmetrigeneral} \end{eqnarray}
with a boundary metric
\begin{equation}
\label{eqmetriboundgen}
ds_0^2 = g_{\mu\nu}^{(0)} dx^\mu dx^\nu = -h(x) dt^2 + \frac{dx^2}{h(x)}.
\end{equation}
In this parametrization, horizons of the boundary metric correspond to
zeros of the function $h(x)$. Minimal curves, starting from the boundary horizon and extending into the bulk, coincide with the bulk
horizon. They are given by the relation
\be
z(x)=4\sqrt{\frac{h(x)}{4\,\mu-h'^2(x)}}.
\label{bulkhor} \ee
Calculating the length of the minimal curve through use of this expression
gives
\be
{\rm Area}(\gamma_A)=
\sqrt{\mu}\int dx \left(\frac{1}{h(x)}+\frac{2\,h''(x)}{4\,\mu-h'^2(x)} \right).
\label{areageneral} \ee
In the vicinity of a zero of $h(x)$ at $x=x_0$,
the first term gives the leading contribution.
Expanding $h(x)$ around this point and using eq. (\ref{bulkhor}), the integration
in eq. (\ref{areageneral}) can be rewritten in terms of $z$ near zero.
Each horizon gives a contribution
\be
{\rm Area}(\gamma_A)=
\frac{2\sqrt{\mu}}{|h'(x_0)|}\int_\epsilon \frac{dz}{z}=
\frac{2\sqrt{\mu}}{|h'(x_0)|}\log\left( \frac{1}{\epsilon} \right).
\label{areageneralres} \ee
For a de Sitter boundary, the minimal curves approaches two horizons, while
$|h'(x_0)|=2H$, resulting in eq. (\ref{areadsbhcutoff}). For a black-hole boundary,
there is only one horizon, while $|h'(x_0)|=2$, resulting in eq. (\ref{areabhbhcutoff}).

As we have already pointed out, the only element that differentiates the final result for
the various metrics that we are considering is the way the divergence near the boundary is regulated.
We assume that the physical cutoff must be imposed on the
bulk coordinate in the Fefferman-Graham parametrization
for every choice of boundary metric. This is consistent with
the framework of holographic renormalization. Even though this
coordinate is always denoted by $z$, setting $z=\epsilon$ implements a
different cutoff procedure each time. This is reflected explicitly in the result of eq. (\ref{areageneralres}).  


We expect that our findings generalize for higher-dimensional AdS spaces with bulk
black holes \cite{marolfds,fischlerds}, as well as for 
similar holographic setups produced by non-trivial bulk potentials \cite{Casalderrey-Solana:2020vls}.

\section*{Acknowledgments}

We would like to thank C. Bachas and G. Pastras for useful discussions.
The research work of D.G. was supported by 
the Hellenic Foundation for Research and Innovation (H.F.R.I.)
and the General Secretariat for Research and Technology (G.S.R.T.), under grant agreement No 2344.
The research work of N.T. was supported by 
the Hellenic Foundation for Research and Innovation (H.F.R.I.) under the 
“First Call for H.F.R.I. Research Projects to support Faculty members and 
Researchers and the procurement of high-cost research equipment grant” 
(Project Number: 824).


\begin{thebibliography}{999}



\bibitem{sorkin}
L.~Bombelli, R.~K.~Koul, J.~Lee and R.~D.~Sorkin,
Phys. Rev. D \textbf{34} (1986), 373-383;
  \\
  M.~Srednicki,
  Phys.\ Rev.\ Lett.\  {\bf 71} (1993) 666
  [hep-th/9303048].

  
\bibitem{jacobson}
 L.~Susskind and J.~Uglum,
  Phys.\ Rev.\ D {\bf 50} (1994) 2700
  [hep-th/9401070];
\\
T.~Jacobson,
[arXiv:gr-qc/9404039 [gr-qc]].

\bibitem{giataganas}
D.~Giataganas and N.~Tetradis,
Phys. Lett. B \textbf{796} (2019), 88-92
[arXiv:1904.13119 [hep-th]].

\bibitem{correctionsds}
N.~Tetradis,
Phys. Lett. B \textbf{807} (2020), 135552
[arXiv:1910.10587 [hep-th]].

\bibitem{adscft1}
J.~M.~Maldacena,
Int. J. Theor. Phys. \textbf{38} (1999), 1113-1133
[arXiv:hep-th/9711200 [hep-th]].

\bibitem{adscft2}
S.~S.~Gubser, I.~R.~Klebanov and A.~M.~Polyakov,
Phys. Lett. B \textbf{428} (1998), 105-114
[arXiv:hep-th/9802109 [hep-th]];
\\
E.~Witten,
Adv. Theor. Math. Phys. \textbf{2} (1998), 253-291
[arXiv:hep-th/9802150 [hep-th]].
  
\bibitem{ryu}
  S.~Ryu and T.~Takayanagi,
  Phys.\ Rev.\ Lett.\  {\bf 96} (2006) 181602
  [hep-th/0603001];
  \\
  V.~E.~Hubeny, M.~Rangamani and T.~Takayanagi,
  JHEP {\bf 0707} (2007) 062
  [arXiv:0705.0016 [hep-th]];
    \\
    T.~Nishioka, S.~Ryu and T.~Takayanagi,
    J.\ Phys.\ A {\bf 42} (2009) 504008
    [arXiv:0905.0932 [hep-th]].

\bibitem{review}
  S.~Ryu and T.~Takayanagi,
  JHEP {\bf 0608} (2006) 045
  [hep-th/0605073];
  \\
  T.~Nishioka, S.~Ryu and T.~Takayanagi,
  J.\ Phys.\ A {\bf 42} (2009) 504008
  [arXiv:0905.0932 [hep-th]].

\bibitem{hawking}
 S.~Hawking, J.~M.~Maldacena and A.~Strominger,
  JHEP {\bf 0105} (2001) 001
  [hep-th/0002145].

\bibitem{rs}  
  L.~Randall and R.~Sundrum,
  Phys.\ Rev.\ Lett.\  {\bf 83} (1999) 3370
  [hep-ph/9905221];
  Phys.\ Rev.\ Lett.\  {\bf 83} (1999) 4690
  [hep-th/9906064].

\bibitem{hol1}
  V.~Balasubramanian and P.~Kraus,
  Commun.\ Math.\ Phys.\  {\bf 208} (1999) 413
  [hep-th/9902121]; 
  \\
  R.~Emparan, C.~V.~Johnson and R.~C.~Myers,
  Phys.\ Rev.\ D {\bf 60} (1999) 104001
  [hep-th/9903238];
\\
  P.~Kraus, F.~Larsen and R.~Siebelink,
  Nucl.\ Phys.\ B {\bf 563} (1999) 259
  [hep-th/9906127].

\bibitem{skenderis}  
  S.~de Haro, S.~N.~Solodukhin and K.~Skenderis,
  Commun.\ Math.\ Phys.\  {\bf 217} (2001) 595
  [hep-th/0002230];
\\
  K.~Skenderis,
  Class.\ Quant.\ Grav.\  {\bf 19} (2002) 5849
  [hep-th/0209067].

\bibitem{marolfds}
D.~Marolf, M.~Rangamani and M.~Van Raamsdonk,
Class. Quant. Grav. \textbf{28} (2011), 105015
[arXiv:1007.3996 [hep-th]].

\bibitem{fischlerds}
W.~Fischler, S.~Kundu and J.~F.~Pedraza,
JHEP \textbf{07} (2014), 021
[arXiv:1311.5519 [hep-th]].
  
\bibitem{probequark}
J.~M.~Maldacena,
Phys. Rev. Lett. \textbf{80} (1998), 4859-4862
[arXiv:hep-th/9803002 [hep-th]];
\\
A.~Brandhuber, N.~Itzhaki, J.~Sonnenschein and S.~Yankielowicz,
Phys. Lett. B \textbf{434} (1998), 36-40
[arXiv:hep-th/9803137 [hep-th]];
JHEP \textbf{06} (1998), 001
[arXiv:hep-th/9803263 [hep-th]].  
  

\bibitem{marolfbh}
V.~E.~Hubeny, D.~Marolf and M.~Rangamani,
Class. Quant. Grav. \textbf{27} (2010), 095015
[arXiv:0908.2270 [hep-th]].
  
  
  
\bibitem{btz}
M.~Banados, C.~Teitelboim and J.~Zanelli,
Phys. Rev. Lett. \textbf{69} (1992), 1849-1851
[arXiv:hep-th/9204099 [hep-th]];
\\
M.~Banados, M.~Henneaux, C.~Teitelboim and J.~Zanelli,
Phys. Rev. D \textbf{48} (1993), 1506-1525
[erratum: Phys. Rev. D \textbf{88} (2013), 069902]
[arXiv:gr-qc/9302012 [gr-qc]].
  




\bibitem{fg}
C.~Fefferman and C.~Robin~Graham, {\em Conformal Invariants}, in 
{\em Elie Cartan et les Math\'ematiques d'aujourd'hui}, Ast\'erisque, 1985,  page~95.



\bibitem{skenderisbh}
K.~Skenderis and S.~N.~Solodukhin,
Phys. Lett. B \textbf{472} (2000), 316-322
doi:10.1016/S0370-2693(99)01467-7
[arXiv:hep-th/9910023 [hep-th]].



\bibitem{papadimitriou}
I.~Papadimitriou and K.~Skenderis,
IRMA Lect. Math. Theor. Phys. \textbf{8} (2005), 73-101
[arXiv:hep-th/0404176 [hep-th]].

  
\bibitem{bekenstein}
  J.~D.~Bekenstein,
  Phys.\ Rev.\ D {\bf 7} (1973) 2333;
  \\
  S.~W.~Hawking,
  Commun.\ Math.\ Phys.\  {\bf 43} (1975) 199
   Erratum: [Commun.\ Math.\ Phys.\  {\bf 46} (1976) 206].
    
\bibitem{Chu:2016pea}
C.~S.~Chu and D.~Giataganas,
Phys. Rev. D \textbf{96} (2017) no.2, 026023
[arXiv:1608.07431 [hep-th]].


\bibitem{Casalderrey-Solana:2020vls}
J.~Casalderrey-Solana, C.~Ecker, D.~Mateos and W.~van der Schee,
[arXiv:2011.08194 [hep-th]].




\end{thebibliography}
\end{document}